# Prediction of topological Hall effect in a driven magnetic domain wall

Kab-Jin Kim[1]★, Masahito Mochizuki[2]★, & Teruo Ono[3,4]★

[1]*Department of Physics, Korea Advanced Institute of Science and Technology, Daejeon 34141, Korea*
[2]*Department of Applied Physics, School of Advanced Science and Engineering, Waseda University, 169-8555, Japan*
[3]*Institute for Chemical Research, Kyoto University, Gokasho, Uji, Kyoto, 611-0011, Japan*
[4]*Center for Spintronics Research Network (CSRN), Graduate School of Engineering Science, Osaka University, Machikaneyama 1-3, Toyonaka, Osaka 560-8531, Japan.*

★Correspondence to: kabjin@kaist.ac.kr, masa_mochizuki@waseda.jp, ono@scl.kyoto-u.ac.jp

**We investigate the possible emergence of topological Hall effect (THE) in a driven magnetic DW. Numerical simulation based on the Landau-Lifshitz-Gilbert-Slonczewski (LLGS) equation shows that the emergent magnetic flux appears when the DW is in a non-equilibrium state. The magnitude of magnetic flux is modulated by Dzyaloshinskii–Moriya interaction (DMI), providing an experimental test of the predicted THE by a voltage controlled DMI modulation. These results indicate that the THE can be observed even in a topologically trivial magnetic DW, and therefore open up new possibility to electrically detect the dynamical spin structure.**

When conduction electrons move in a system with spatially varying magnetization, their spins follow the local magnetization direction if the exchange interaction between the conduction-electron spin and the local magnetic moment is strong enough. In its rest frame, the electrons feel a time-dependent magnetic field and thus pick up a so-called Berry phase that depends on the solid angles subtended by the noncoplanar magnetization directions. This Berry phase is known to induce a novel Hall effect, which was termed topological Hall effect (THE) as it depends solely on the topology of spin textures[1-4]. The THE has recently been observed in several systems where a skyrmion lattice exists due to the Dzyaloshinskii–Moriya interaction (DMI)[5-13].

In contrast to the skyrmions, magnetic domain walls (DWs) have topologically trivial spin textures[14]. Hence, it has been considered that the THE would be absent for the static magnetic DW. In its nonequilibrium state, however, the DW can have non-trivial spin textures such as vertical Bloch lines[15], and therefore may induce a THE. Based on this naïve expectation, here we investigate possible emergence of THE in a driven magnetic DW by using micromagnetic simulation. Numerical simulations show that non-trivial spin textures appear when the DW moves due to the formation and propagation of vertical Bloch lines. This non-trivial spin textures are found to have finite topological charges which induces emergent magnetic flux, implying that the THE can be observed in a driven magnetic DW. We find that magnitude of the emergent magnetic flux can be modulated by the DMI, providing an experimental test of the predicted THE by a voltage controlled DMI modulation[16-17]. This result suggests that the THE can be observed even in a magnetic DW motion, and therefore open up a new opportunity to electrically detect the dynamical topological spin structures.

For this study, we perform micromagnetic simulations using the Landau-Lifshitz-Gilbert-Slonczewski equation which describes magnetization dynamics in the presence of spin-polarized electric currents as well as an external magnetic field $H$. We assume a typical ferromagnetic nanowire with a perpendicular magnetic anisotropy on top of a heavy-metal layer in which the field- and current-driven DW motion under the influence of DMI can easily be realized[15, 18-19]. To mimic the experimental situation, we further assume a sample geometry with a Hall cross structures fabricated to measure the Hall resistivity as shown in Fig. 1(a). We examine a 5 μm-wide and 1.2 nm-thick wire with 100nm-wide Hall bar. For the simulations, we divided the wire into equivalent rectangular cells of size $2\times2\times1.2$ nm$^3$, in which the magnetizations are assumed to be uniform in each cell. We assume the exchange stiffness $A=1.0\times10^{-11}$ [Jm$^{-1}$], the DM parameter $D = 0.5\times10^{-3}$ [Jm$^{-2}$], saturation magnetization $M_s=3.0\times10^5$ [Am$^{-1}$], and the uniaxial magnetic anisotropy $K_U=2.0\times10^5$ [Jm$^{-3}$] according to the parameter evaluations for typical ferromagnetic nanowires in Ref. [18]. The Gilbert-damping coefficient is set to be α=0.2. With this setup, we study the magnetic-field-driven motion of the DW. To examine more realistic experimental situation, we consider the influence of electric currents injected to measure the Hall voltage by incorporating the spin-transfer torque term and the nonadiabatic term in the Landau-Lifshitz-Gilbert-Slonczewski equation. The electric current density $j$ is set to be $3.0\times10^{10}$ A/m$^2$ according to the experimental setup in Ref. [18]. The spin-polarization of the electric current $p$ and the strength of the nonadiabatic term β are set to be $p=0.5$ and $β=2α=0.4$, respectively[18].

By numerically solving the equation using the fourth-order Runge-Kutta method, we trace temporal evolutions of magnetizations in the ferromagnetic wire under the external magnetic field $\boldsymbol{H}=(H_x, 0, H_z)$, which is inclined from the vertical direction towards a longitudinal direction along the magnetic wire. We fixed the out-of-plane field component $H_z =150$ mT, which is a typical value to drive DW[15]. Using the obtained spatio-temporal magnetization configuration et each time step, we calculate total topological charge of the noncollinear magnetization structure emerging at the domain-wall region, which corresponds to a sum of the scalar spin chiralities or equivalently the solid angles spanned by three neighboring magnetization vectors $\boldsymbol{m}$ divided by $4\pi$. Here, the sum of the scalar spin chiralities $\boldsymbol{S}$ is given by

$$\boldsymbol{S} = \int d\boldsymbol{r} \left(\frac{\partial \boldsymbol{m}}{\partial x} \times \frac{\partial \boldsymbol{m}}{\partial y}\right) \cdot \boldsymbol{m}, \tag{1}$$

where $m$ is the normalized magnetization vector at the position $r$. A topological charge $Q=\mp 1$ works as one magnetic-flux quantum of $h/2e$ acting on the conduction-electron spins, which is known to induce the THE.

Figure 1(b) shows a typical magnetization profile of driven magnetic DW. To mimic the realistic experimental situation, we start with a nonuniform DW structure which contains some Bloch lines as defects. We note that the Bloch line defects are necessarily generated by impurity pinning effects and thermal excitation in actual experiments. We trace the time profiles of emergent magnetic flux observed in a small area, which is assigned to the Hall cross area in the actual experimental system [gray region in Fig. 1(b)]. Figure 1(c) shows a simulated time profile of the emergent magnetic flux when $H$ is normal to the nanowire plane with $H_x=0$. The effective magnetic field corresponding to the emergent magnetic flux is also given in right axis of Fig. 1(c) to provide more realistic value of magnetic flux. We find that a large number of magnetic fluxes appear when the DW passes through the Hall cross area, indicating that the large THE can be observed in the DW motion.

To elucidate an origin of the emergent magnetic fluxes, we took a snapshot of the magnetization configuration of the domain wall at a selected time in the simulation. Figure 2(a) shows a global view of the domain wall by mapping the $m_z$ component of magnetization vectors with colors, while Fig. 2(b) shows its magnified view. We find that the moving DW is meandering and pinned at some points due to the vertical Bloch lines (VBLs), because the VBLs generally play as defects for DW propagation[15]. As a result, the DW shows an intermittent motion, that is, portions between the pinned points become convex leftward and grow to be depinned. In the course of this stick-slip type motion, the pinned points become swelled to form skyrmion-like magnetic structures eventually as shown in Figs. 2(b) and 2(c). A magnified view of this skyrmion-like magnetic structure is shown in Fig. 2(d). The colors in Fig. 2(c) and 2(d) indicate spatial distributions of the local scalar spin chiralities. Here, the sign of the scalar spin chirality is determined by the sign of the external magnetic field applied to drive the domain wall. When the external magnetic field is applied in the positive z direction, magnetizations at the periphery of the created skyrmion-like spin texture must be oriented along the positive z direction to maximize an energy gain from the Zeeman interaction, while the magnetization at the center of the spin texture inevitably points in the opposite direction. This magnetization configuration eventually gives rise to a negative sign of the scalar spin chirality. The observed negative spin chiralities correspond to positive emergent magnetic field for conduction electron spins, which inevitably induces the THE.

We next examine the effect of DMI on the THE. It has been known that the THE is closely related with the DMI, since the DMI generates the chiral spin textures with finite topological charges that is essential for THE. Figure 3 shows the time profiles of emergent magnetic flux (or magnetic field) for several strengths of the DMI. Two features are observed. First, the DW arrival time becomes shorter for a stronger DMI, meaning that the DW velocity increases with increasing DMI. This tendency is in good agreement with the recent experimental observations in Ref. [15], suggesting that our micromagnetic simulation well reproduces the experiments. Second feature is that the magnitude of magnetic flux is modulated by the DMI: the stronger the DMI, the larger the magnetic flux. This implies that the larger THE can be obtained in samples with a strong DMI. Importantly, we also found in Fig. 3 that the emergent magnetic flux almost vanishes when the DMI is absent ($D$=0). This is because the DMI generally induces a drastic DW bending due to the formation and collision of the vertical Bloch lines [15]. The collisions of vertical Bloch lines render the DW meander more significantly, which results in a larger THE.

Lastly, we discuss possible experimental verification of the predicted THE. The THE predicted here is a general phenomenon expected in driven DWs and thus should be observed experimentally. It was recently succeeded to measure the real-time DW motion by monitoring anomalous Hall effect (AHE) resistance change[15,20-22]. Therefore, it is possible to observe the THE in a non-equilibrium DW state. Based on our simulations, we predict that the THE can be observed as a superposed Hall-resistance signal to that of the AHE. More concretely, a small peak should appear in the Hall-resistance signal when the DW passes through the Hall-cross area or when the AHE signal changes its sign. A rough estimation based on the Hall resistivity measurement of metallic ferromagnets gives an expected THE voltage of the order of microvolt, enough large to be measured using a commercial nano-voltmeter. Once the peak is observed, one can verify that it is indeed of the THE origin by modulating the DMI by electric field, because the DMI can be largely modulated by the voltage application[16,17]. Therefore, the dependence on the electric field gives a way to test the predicted THE in the experiment.

In summary, we investigated the possible emergence of THE in a magnetic DW motion. Although the magnetic DW itself has a topologically trivial spin textures in an equilibrium state, we find that the magnetic DW can generate THE in a non-equilibrium state. When the DW moves, it starts to bend and form a skyrmion-like spin structures. This non-trivial spin textures generate emergent magnetic fluxes, which give a THE. The magnitude of the emergent magnetic fluxes is found to be modulated by DMI, providing a way to verify the THE experimentally by voltage controlled DMI modulation. Our work will provide a way to create and observe topologically non-trivial spin structures in a dynamical regime, and therefore paves the way for future topology-based spintronic applications.

This work was partly supported by JSPS KAKENHI Grant Numbers 15H05702, 26870300, 26870304, 26103002, 25220604, 2604316 Collaborative Research Program of the Institute for Chemical Research, Kyoto University, the Cooperative Research Project Program of the Research Institute of Electrical Communication, Tohoku University, and R & D project for ICT Key Technology of MEXT from the Japan Society for the Promotion of Science (JSPS). KJK was supported by the National Research Foundation of Korea (NRF) grant funded by the Korea government (MSIP) (No. 2017R1C1B2009686, NRF-2016R1A5A1008184). MM thanks supports by JSPS KAKENHI (Grant No. 17H02924), Waseda University Grant for Special Research Projects (Project No. 2018K-257), and JST PRESTO (Grant No. JPMJPR132A).

## Figure Legends

**Figure 1| (a)** Schematic illustration of the experimental situation considered for the micromagnetic simulation. **(b)** Snapshots of the domain wall before its arriving at the Hall cross area (the gray area). **(c)** Simulated time profile of the emergent magnetic flux observed in the Hall cross area. The right vertical axis shows the emergent magnetic field (the magnetic-flux density) calculated by dividing the magnetic flux number by the approximate area of the domain wall. Here the width (area) of the domain wall is roughly estimated as 0.2 μm (1 μm$^2$).

**Figure 2| (a)** Global view of the moving domain wall with mapping of the magnetization $m_z$ component. **(b)** Magnified view of Fig. 2(a). **(c)** Spatial distribution of local scalar spin chiralities (colors) and in-plane magnetization components (arrows). A skyrmion-like spin structure appearing in the domain wall is indicated by red (solid) circle. **(d)** Magnified view of the skyrmion-like spin texture in Fig. 2(c).

**Figure 3|** Time profiles of the emergent magnetic flux observed in the Hall cross area for several strengths of the DMI.

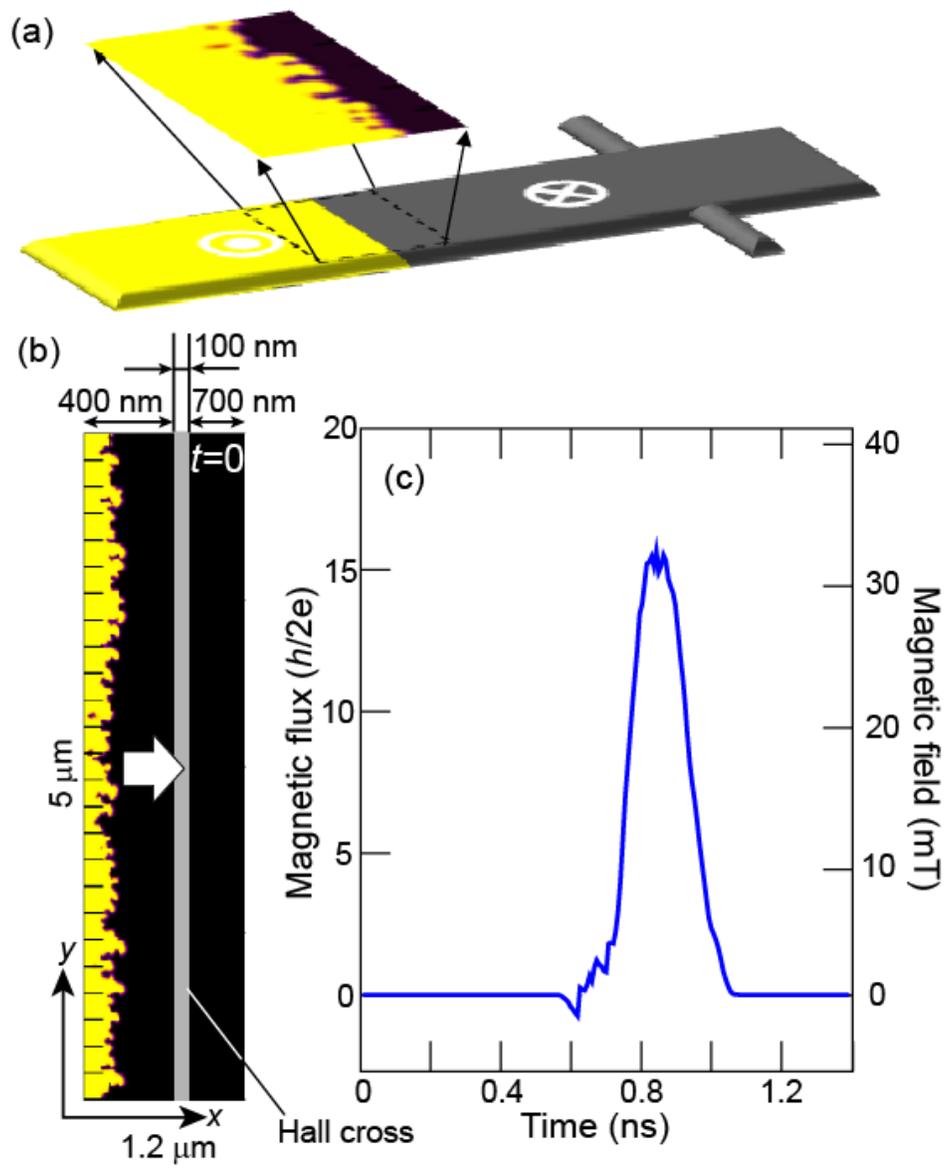

Fig. 1

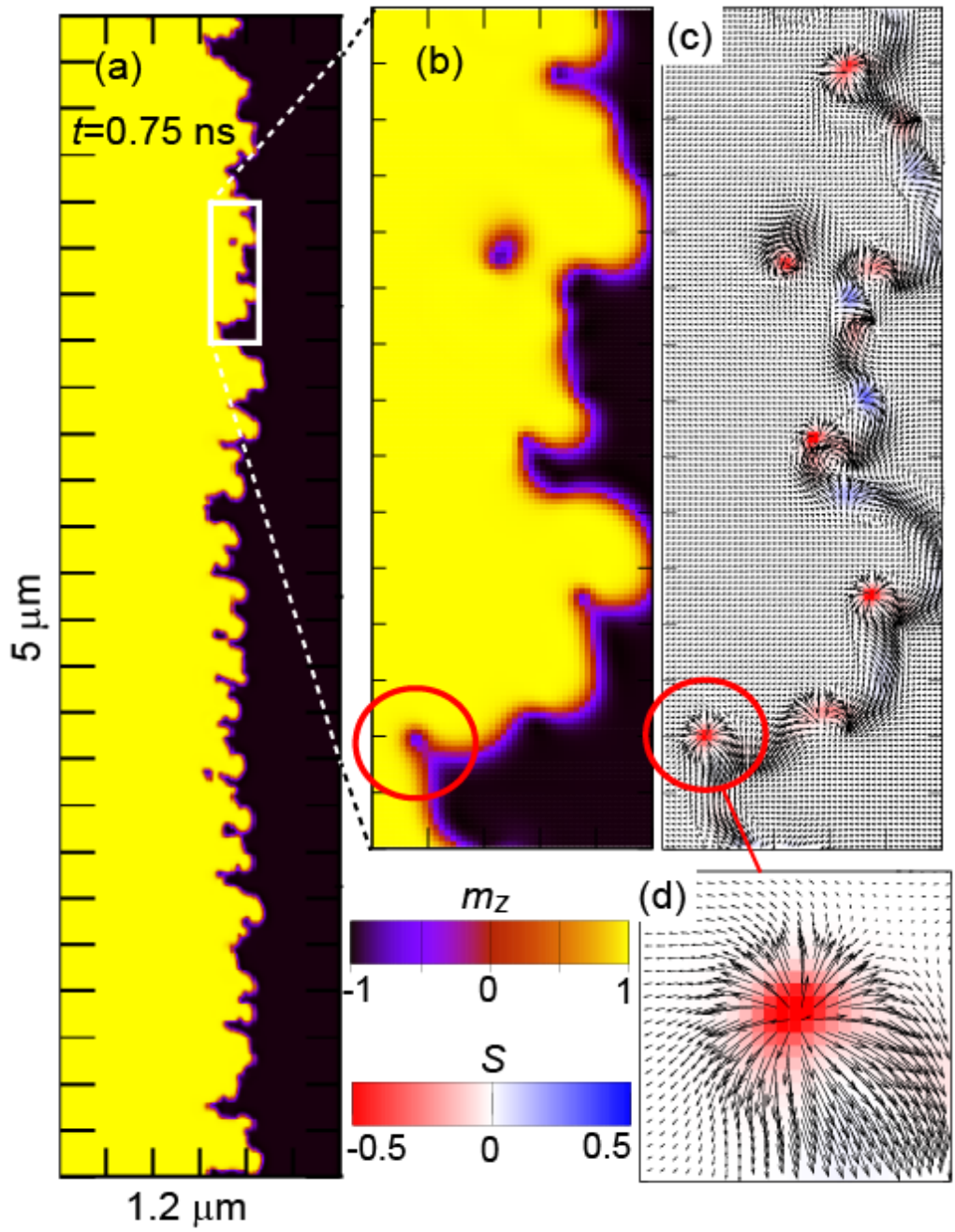

Fig. 2

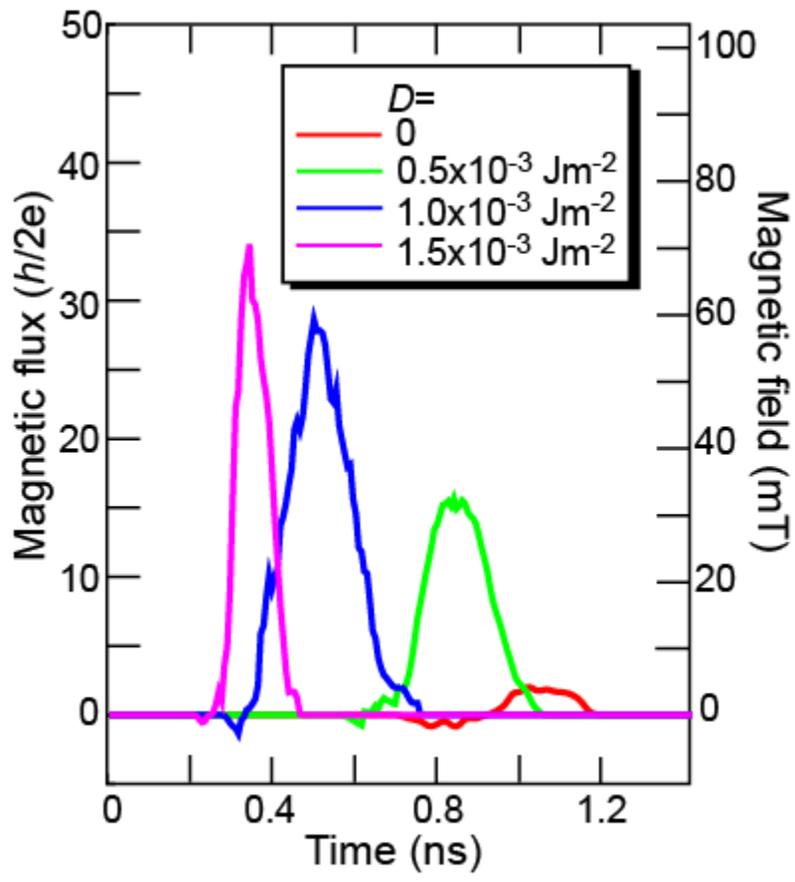

Fig. 3